\documentclass[twocolumn]{aastex63}
\usepackage{natbib}
\usepackage{multirow}
\usepackage{amsmath}
\usepackage{xcolor}
\usepackage{CJK}
\usepackage{graphicx}
\usepackage{float}
\usepackage{soul}
\usepackage{color, subfigure, lineno}

\def\nod{\nodata}

\shorttitle{VODKA: VLA Radio Classification}
\shortauthors{Baltaci et al.}

\begin{document}
\begin{CJK*}{UTF8}{gbsn}

\title{Varstrometry for Off-nucleus and Dual sub-Kpc AGN (VODKA): Radio Classification of High-Redshift Dual AGN Candidates with the Very Large Array}

\correspondingauthor{Sude Baltaci}
\email{baltaci2@illinois.edu}

\author[0009-0006-9161-0462]{Sude Baltaci}
\affiliation{Department of Astronomy, University of Illinois at Urbana-Champaign, Urbana, IL 61801, USA}

\author[0000-0001-7681-9213]{Arran C. Gross}
\affiliation{Department of Astronomy, University of Illinois at Urbana-Champaign, Urbana, IL 61801, USA}

\author[0000-0003-0049-5210]{Xin Liu}
\affiliation{Department of Astronomy, University of Illinois at Urbana-Champaign, Urbana, IL 61801, USA}

\affiliation{National Center for Supercomputing Applications, University of Illinois at Urbana-Champaign, Urbana, IL 61801, USA}

\author[0000-0001-6100-6869]{Nadia Zakamska}
\affiliation{William H. Miller III Department of Physics and Astronomy, Johns Hopkins University, Baltimore, MD 21210, USA}

\author[0000-0003-1659-7035]{Yue Shen}
\affiliation{Department of Astronomy, University of Illinois at Urbana-Champaign, Urbana, IL 61801, USA}
\affiliation{National Center for Supercomputing Applications, University of Illinois at Urbana-Champaign, Urbana, IL 61801, USA}

\author[0000-0003-3673-7314]{Mingrui Liu}
\affiliation{William H. Miller III Department of Physics and Astronomy, Johns Hopkins University, Baltimore, MD 21210, USA}

\begin{abstract}
Dual active galactic nuclei (dual AGNs) are pairs of simultaneously accreting supermassive black holes in merging galaxies. We investigate dual AGNs to understand whether merger-induced accretion is a significant growth mechanism for supermassive black holes. Searching for such systems is favorable at close separations and high redshift (Cosmic Noon, \(z \sim 2 \)) due to the expected combination of high galaxy merger rate and peak AGN activity which characterize this era of the Universe. The sample of nine dual AGN candidates is selected based on resolved optical dual detections with Gaia, whose angular separations are less than 1\arcsec \ and redshifts range between 1.5 and 2.8. Each pair is spatially coincident with a Sloan Digital Sky Survey quasar primary target. We aim to classify the secondary targets and other components in the radio regime using 2-band Very Large Array imaging (C and Ku-bands) to test for dual AGN presence. We identify two dual AGNs and three quadruply imaged gravitational lens AGNs, two out of which show evidence of radio flux anomaly. The two new dual AGNs add to the limited census of confirmed kpc-scale pairs at Cosmic Noon, while the radio flux anomalies in J0911+0550 and J1118+0745 provide independent evidence for substructure in their lensing potentials, consistent with microlensing on compact AGN emission regions. Besides one confirmed AGN-star pair, three candidates remain unclassified due to lack of radio detection for one or two components.
\end{abstract}

\keywords{black hole physics --- galaxies: active --- quasars: general --- surveys}


\section{Introduction}\label{sec:intro}

Supermassive black holes (SMBHs) with masses ranging from \(10^6 M_\odot\) to \(10^{10} M_\odot\) \citep{McConnell2013} are found in the centers of most, if not all, galaxies \citep{Kormendy1995}. They are predicted to have grown primarily by accretion of surrounding materials \citep{Yu2002} and mergers with smaller black holes \citep{DiMatteo2005}. These SMBHs evolve with their host galaxies and their growth appears to be linked to the star formation history of the galaxy. This is most clearly shown by the correlation between the mass of the SMBH and the velocity dispersion of the bulge of the host galaxy across many orders of magnitude \citep{Kormendy2013}. The relation between the bulge mass and SMBH mass equally suggests coevolution of SMBHs and at least the inner regions of their host galaxies \citep{Kormendy2013}. 

An accreting SMBH which starts emitting light by the disk that get reprocessed into many wavelengths by the surrounding structures is called an active galactic nucleus (AGN) \citep{Hernquist1995}.
Quasars are the most optically luminous AGNs with bolometric luminosities \(L_{bol}  \gtrsim 10^{45} \ \mathrm{erg\ s^{-1}}\). Some AGNs are bright in radio band due to synchrotron-emitting jets.

Simulations predict that galaxy mergers can be a reason for SMBHs to become active and start emitting across the electromagnetic spectrum \citep{VanWassenhove2012, Rosas-Guevara2019}. Mergers redirect the gas in the galaxies, allowing it to flow inwards to the central black holes \citep{Hernquist1995, DiMatteo2005}. Hence, we expect mergers to be an important mechanism for triggering AGNs, including their most luminous periods of growth. The time period known as the Cosmic Noon corresponds to a redshift around 2, which is when the Universe was around 3 billion years old. This era is marked by intense star formation, black hole growth and mergers, making it favorable for searching for dual AGNs where it is common to find instances of at least one AGN in a galaxy merger \citep{Richards2006}. Thus, we aim to answer whether galaxy mergers trigger dual AGN at Cosmic Noon. 

Our sample is a subset of the sample presented in \citet{Shen2023}, which consists of 50 dual AGN candidates separated by \(0\farcs2 - 0\farcs9\). This sample is assembled from the Sloan Digital Sky Survey Data Release 16 (SDSS DR16) \citep{Lyke2020} with each pair containing at least one spectroscopically confirmed Type I quasar based on the best-fit coordinates in SDSS and principal component analysis (PCA) to rule out blended stellar interlopers. PCA is a method based on \(\chi^2\) minimization which decomposes the observed blended spectrum into the underlying spectra of the multiple contributing objects to find the best fit among different model spectra.

\citet{Shen2023} picked out sources with \(z \geq 1.5\) in order to make emissions from the host galaxy in Gaia bandpass negligible. This leaves predominantly quasar emission in the image since galaxy spectra in the ultraviolet range taper off in the rest-frame. Out of these quasars, the ones with the possibility of being duals were chosen by searching in Gaia Early Data Release 3 (EDR3) \citep{Gaia2023} for sources in a 3\arcsec \ radius circle around the SDSS sources. The pairs were searched using Gaia EDR3 as SDSS fiber size encloses both suspected targets in a pair, making it difficult to determine which target is the source of the quasar emission. Only targets with magnitudes brighter than Gaia \textit{G}-band magnitude \(G\leq 20.25\) out of the matched Gaia sources were included in the sample to reduce uncertainty. 

Between the two sources in a given pair, the brighter source as observed by Gaia is called the primary. Out of the 50 dual AGN candidates, 30 with the smallest separations were chosen for further analysis in radio band. We investigate the radio emission of 9 targets which received 2-band National Science Foundation's Karl G. Jansky Very Large Array (VLA) imaging. Throughout this project, we assume a flat Lambda cold dark matter (\(\Lambda\)CDM) cosmology with values of \(\Omega_{\Lambda} = 0.7\), \(\Omega_{m} = 0.3\), and \(H_{0} = 70 \ \mathrm{km \ s^{-1} \ Mpc^{-1}}\).


\section{Methods}\label{sec:method}
As radio waves are less sensitive to dust obscuration compared to optical wavelengths, we use radio data to analyze the sources. The VLA data (Table \ref{tab:VLA_obs}) was obtained in the A-configuration between October 15, 2024 and January 26, 2025 during the 2024B semester (Program VLA/24B-361; PI: A. Gross). It is first run through the VLA pipeline using the Common Astronomy Software Applications (CASA) \citep{CASATeam2022}. This automated step is crucial in calibrating the raw telescope output, flagging corrupt or incomplete data and performing an initial cleaning for weather conditions, electronics issues, and radio frequency interference. We then reduce the calibrated datasets using the CASA \texttt{tclean} algorithm and model them using the CASA \texttt{imfit} algorithm. 

We use the \texttt{tclean} algorithm twice for each observation in each band to remove interference from bright point sources near the target. This is done to ensure that the rms noise is down to a target value: 4.3 \(\mathrm{\mu Jy \ beam^{-1}}\) for C-band and 4.4 \(\mathrm{\mu Jy \ beam^{-1}}\) for Ku-band. These rms noise values are chosen based on the faintest observed AGN luminosities in previous studies with a flux density of \(\sim\)15 \(\mathrm{\mu Jy}\) at Cosmic Noon \citep{Gross2023}. The 3\(\sigma\) flux limit corresponds to the rest-frame luminosity of \(8.4 \times10^{29} \mathrm{erg \ s^{-1} \ Hz^{-1}}\) in C-band and \(8.5 \times10^{29} \mathrm{erg \ s^{-1} \ Hz^{-1}}\) in Ku-band, falling in the radio-quiet AGN range of \(L_{5 \mathrm{GHz}} \sim 10^{29} - 10^{31} \ \mathrm{erg \ s^{-1} \ Hz^{-1}}\). Therefore, we ensure a significant portion of radio-quiet AGNs will be observed.

\begin{deluxetable}{lcccccr}
\tabletypesize{\scriptsize}
\tablewidth{0.4\textwidth}
\tablecaption{VLA Observations
\label{tab:VLA_obs}}
\tablehead{ 
\colhead{Target} & \colhead{UT} & \colhead{IT} & \colhead{rms} & \colhead{Beam} & \colhead{PA} & \colhead{Calibrator} \\
\colhead{J200}  & \colhead{date} & \colhead{min} & \colhead{$\mu$Jy/beam} & \colhead{(arcsec)} & \colhead{(deg)} & \colhead{} \\
\colhead{(1)} & \colhead{(2)} & \colhead{(3)} & \colhead{(4)} & \colhead{(5)} & \colhead{(6)} & \colhead{(7)}
}
\startdata 
\multicolumn{6}{c}{C-Band ($\nu$ = 6 GHz)} \\ 
J0041-0203 & 2024-10-23 & 55 & 0.0055 & 0.64$\times$0.27 & -51  & 3C48 \\ 
J0743+2209 & 2024-11-05 & 55 & 0.0041 & 0.32$\times$0.28 & +53 & 3C138 \\ 
J0813+2545 & 2024-11-20 & 55 & 0.0050 & 0.29$\times$0.27 & +7 & 3C138 \\ 
J0911+0550 & 2024-10-15 & 55 & 0.0052 & 0.41$\times$0.28 & -49 & 3C138\\ 
J0940+3346 & 2025-01-26 & 55 & 0.0042 & 0.44$\times$0.29 & -80 & 3C138 \\ 
J1010+5705 & 2025-01-25 & 55 & 0.0048 & 0.46$\times$0.27 & -88 & 3C48\\ 
J1118+0745 & 2025-01-10 & 55 & 0.0054 & 0.34$\times$0.28 & +22 & 3C286\\ 
J2246+0344 & 2024-10-16 & 55 & 0.0049 & 0.33$\times$0.28 & -12 & 3C48\\
J2333+2854 & 2024-12-25 & 55 & 0.0052 & 0.85$\times$0.29 & +61 & 3C48\\ 
\multicolumn{6}{c}{Ku-Band ($\nu$ = 15 GHz)} \\ 
J0041-0203 & 2024-10-15 & 65 & 0.0051 & 0.23$\times$0.13 & -44  & 3C48 \\ 
J0743+2209 & 2024-10-28 & 65 & 0.0028 & 0.15$\times$0.12 & -83 & 3C138\\ 
J0813+2545 & 2024-10-27 & 65 & 0.0030 & 0.15$\times$0.12 & +46 & 3C138\\ 
J0911+0550 & 2024-10-25 & 65 & 0.0029 & 0.21$\times$0.12 & -59 & 3C1383\\ 
J0940+3346 & 2024-12-31 & 65 & 0.0031 & 0.19$\times$0.12 & -78 & 3C138\\ 
J1010+5705 & 2024-11-30 & 65 & 0.0046 & 0.15$\times$0.11 & +67 & 3C48\\ 
J1118+0745 & 2025-01-23 & 65 & 0.0043 & 0.13$\times$0.11 & +10 & 3C286\\ 
J2246+0344 & 2024-10-22 & 65 & 0.0052 & 0.14$\times$0.12 & -23 & 3C48\\ 
J2333+2854 & 2025-01-08 & 65 & 0.0048 & 0.20$\times$0.12 & +66  & 3C48 \\ 
\enddata
\tablecomments{ 
(1) Target Designation;
(2) UT date of observation (Y-M-D);
(3) Integration time for the observation, encompassing the kinematic pair;
(4) rms noise for the cleaned image;
(5) Restoring beam size (maj$\times$min axes);
(6) Restoring beam position angle (degrees East of North)
(7) VLA flux calibrator for the target.
}
\end{deluxetable}

First run of \texttt{tclean} is performed to process the measurement set into a format where interfering sources can be identified visually and removed. These sources produce bright ``sidelobe" patterns that may intersect with the target of interest, even from several arcminutes separation. To reduce computation time, we manually specify smaller sub-frames to locally compute the models for these interfering sources, which are then applied to the full image.

The second run of \texttt{tclean} involves more iterations as the algorithm runs until it reaches 3000 iterations or an rms noise of 4.3 \(\mathrm{\mu Jy \ beam^{-1}}\) for C-band and 4.4 \(\mathrm{\mu Jy \ beam^{-1}}\) for Ku-band, whichever happens first. For all of our sources, we reached the target noise level first. We set the pixel size to reach the highest spatial resolution possible while still sampling with 3-5 pixels along the minor axis of the beam. For C-band, pixel size is set to 0\farcs03 for 640 pixels. For Ku-band, pixel size is set to 0\farcs01 for 1000 pixels. In both cases, the weighting method is \texttt{briggs} with \texttt{robust} = 0.5, which achieves a balance between low surface brightness emission without boosting noise excessively. The deconvolver is set to \texttt{clark} \citep{Clark1980} and the gridder is set to \texttt{standard} \citep{CASATeam2022}. 

We fit our sources to elliptical Gaussian models using the CASA \texttt{imfit} algorithm in order to obtain their characteristics, such as integrated flux and angular size. Based on the visual inspection of the \texttt{tclean}-processed image, the coordinates, shapes and sizes of the detected sources are used to set the \texttt{estimates} variable of \texttt{imfit} as initial models. The model image is constructed with iterative amplitude measurements to minimize the variances of the residuals for the best-fit sources \citep{Condon1997}. This algorithm returns the best-fit model image for the sources and their integrated fluxes. However, it is still possible to have targets with non-Gaussian morphologies such as extended jets, for which we visually inspect the images and add additional model components for better fits. We obtain the average integrated flux density across all significant detections to be \(44.6 \ \mathrm{\mu Jy}\), which suggests a sample of mostly faint sources. The targets are on average compact sources, rather than extended, which are unresolved at the scale of the corresponding clean beam. Their properties and classifications, and the details of their fit models are reported in Tables \ref{tab:properties} and \ref{tab:VLA}, respectively.


\section{Results}\label{sec:result}

\begin{deluxetable}{lcccccr}
\tablewidth{0.4\textwidth}
\tablecaption{Target Properties
\label{tab:properties}}
\tablehead{ 
\colhead{Target} & \colhead{\(z\)} & \colhead{Angular Distance} & \colhead{Classification} \\
\colhead{J200} & \colhead{} & \colhead{(arcsec)} & \colhead{} \\
\colhead{(1)} & \colhead{(2)} & \colhead{(3)} & \colhead{(4)}}
\startdata 
J0041-0203 & 1.849$\pm$0.0024 & 0.573 & Unclassified \\ 
J0743+2209 &  2.065$\pm$0.0044 & 0.229 & AGN-Star \\ 
J0813+2545 & 1.510$\pm$0.0019 & 0.817 & Quad Lens \\ 
J0911+0550 & 2.797$\pm$0.0053 & 0.487 & Quad Lens \\ 
J0940+3346 & 1.782$\pm$0.0035 & 0.712 & Dual AGN \\ 
J1010+5705 & 1.971$\pm$0.0029 & 0.827 & Unclassified \\ 
J1118+0745 & 1.738$\pm$0.0025 & 0.475 & Quad Lens \\ 
J2246+0344 & 2.084$\pm$0.0028 & 0.890 & Unclassified \\
J2333+2854 & 1.674$\pm$0.0024 & 0.779 & Dual AGN \\ 
\enddata
\tablecomments{ 
(1) Target Designation;
(2) Redshift and associated error;
(3) Angular distance between the target pair as it was measured by Gaia during sample selection;
(4) Final classification of the target. Unclassified refers to candidates whose primary detections are confirmed to be AGNs with unclassified secondary targets.
}
\end{deluxetable}

\begin{figure*}[h!]
     \centering
        \includegraphics[width = 0.85\textwidth, height = 1.2\textwidth]{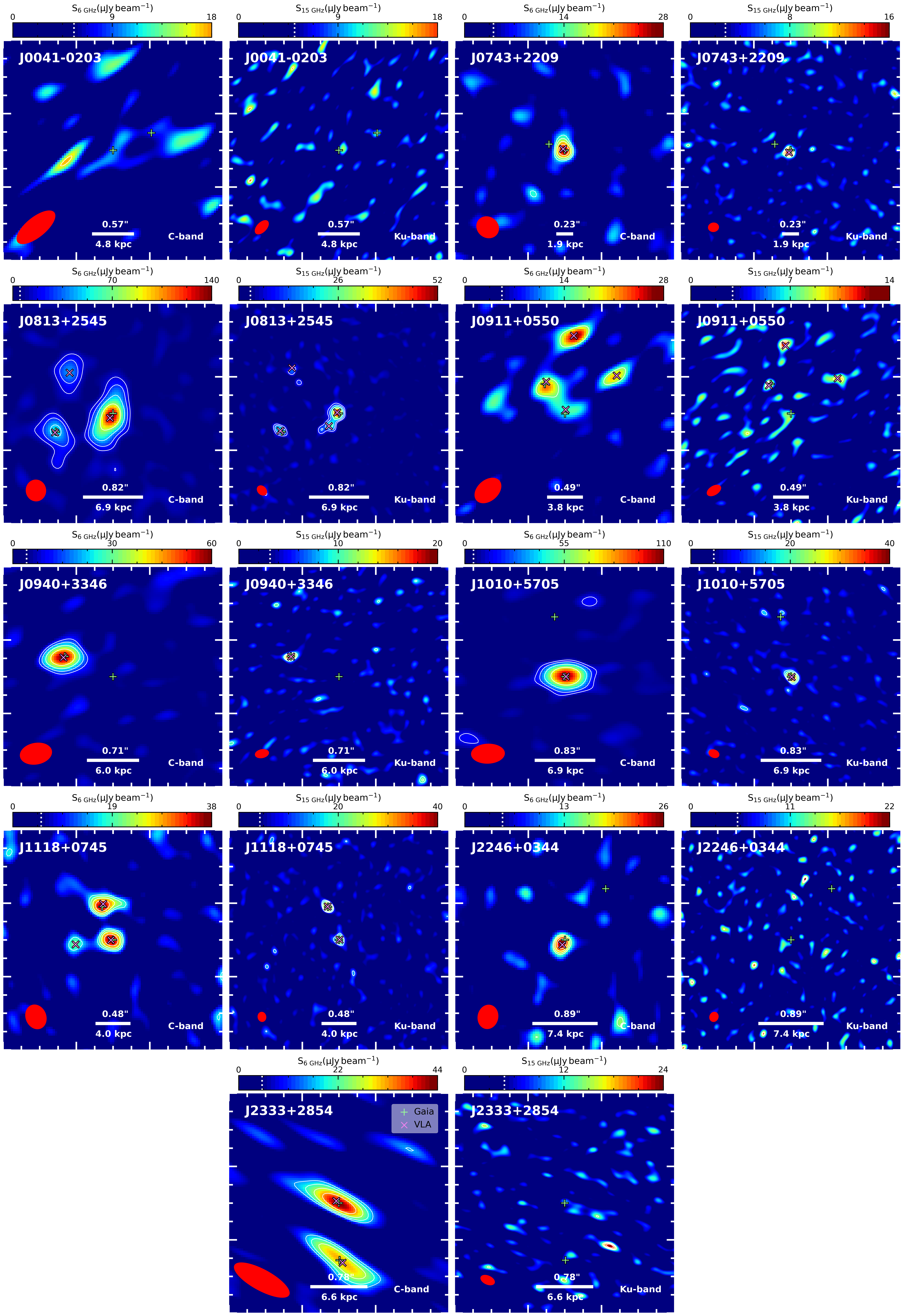}
        \caption{VLA 2-band imaging. Target designation is given in the top left and VLA band is noted in the bottom right of each panel. Panels are centered on the coordinate of the brighter primary source, based on the Gaia detections from \citet{Shen2023} (positions denoted with a green $+$). VLA detections (when present) are denoted with a pink $\times$ (except for J0911+0550 primary, which is marked for completeness despite being a non-significant detection). Major tick marks are spaced 1\arcsec \ apart. Scale bars give the separation between the Gaia positions of the two sources in a given pair. For each panel, the colorbar starts at the rms value (denoted with a dotted line). Contours begin at 3$\times$rms and continue exponentially to the peak value. The restoring beam is shown with a red ellipse in the bottom left corner of each panel.}
     \label{fig:VLA1}
\end{figure*}

The observed images for each target in both VLA bands are shown in Figure \ref{fig:VLA1}.

\begin{itemize}
    \item J0041-0203 showed no radio detection in either VLA band.
    \item J0743+2209 has detections for the primary in both VLA bands, but no detections for the secondary.
    \item J0813+2545 has three detections in C-band, and four detections in Ku-band. The primary and the quaternary detections are blended in C-band. The relative positions of the detections are similar in both bands.
    \item J0911+0550 shows three detections in C-band and three detections in Ku-band. The primary is not a significant detection in either band.
    \item J0940+3346 shows only the detection of the secondary in both bands, and no detection of the primary in either band.
    \item J1010+5705 shows only the detection of the primary in both bands.
    \item J1118+0745 shows the detections of both the primary and the secondary in both bands, with the detection of a third source in C-band.
    \item J2246+0344 shows only the detection of the primary in C-band and no detection of either target in Ku-band.
    \item J2333+2854 shows dual detections in C-band and no detection in Ku-band.
\end{itemize}

\begin{deluxetable*}{lccrrrrrr}
\tabletypesize{\small} \tablewidth{5pt}
\tabletypesize{\scriptsize}
\tablewidth{\textwidth}
\tablecaption{Properties of Radio Detections
\label{tab:VLA}}
\tablehead{ 
\colhead{Target} & \colhead{Band}&  \colhead{Position}&\colhead{$S^{\rm peak}_{\nu}$} & \colhead{$S^{\rm int}_{\nu}$}   & \colhead{$\theta_{maj}$} & \colhead{$\theta_{min}$} & \colhead{PA} & \colhead{$\alpha$}\\
\colhead{J2000}  & \colhead{}& \colhead{J2000}& \colhead{(\(\mathrm{mJy}\)/beam)} & \colhead{(\(\mathrm{mJy}\))}  & \colhead{(marcsec)} & \colhead{(marcsec)} & \colhead{(deg)} & \colhead{(6$-$15 GHz)} \\
\colhead{(1)} & \colhead{(2)} & \colhead{(3)} & \colhead{(4)} & \colhead{(5)} & \colhead{(6)} & \colhead{(7)} & \colhead{(8)} & \colhead{(9)}}
\startdata 
J0743+2209 (1) & C & 07:43:20.91+022.09.45.0 & 0.0249 $\pm$ 0.0039 & 0.0263 $\pm$ 0.0073 & $\leq$409.0$\pm$84.0 & $\leq$229.0$\pm$27.0 & 169.6$\pm$7.8 & \(-0.361\pm0.141\)\\  \vspace{0.20cm} 
J0743+2209 (1) & Ku & 07:43:20.91+022.09.45.0 & 0.0167 $\pm$ 0.0029 & 0.0189 $\pm$ 0.0054 & $\leq$146.0$\pm$26.0 & $\leq$139.0$\pm$24.0 & 128.0$\pm$150.0 & \nod \\ 
J0813+2545 (1,4) & C & 08:13:31.27+025.45.3.0 & 0.1231 $\pm$ 0.0046 & 0.2320 $\pm$ 0.0130 & 455.0$\pm$31.0 & 33.0$\pm$45.0 & 158.7$\pm$2.0 & \nod \\ 
J0813+2545 (2) & C & 08:13:31.33+025.45.2.8 & 0.0490 $\pm$ 0.0049 & 0.0830 $\pm$ 0.0120 & 244.0$\pm$73.0 & 229.0$\pm$92.0 & 31.0$\pm$89.0 & \nod \\ 
J0813+2545 (3) & C & 08:13:31.32+025.45.3.7 & 0.0366 $\pm$ 0.0049 & 0.0760 $\pm$ 0.0140 & 329.0$\pm$96.0 & 255.0$\pm$100.0 & 144.0$\pm$47.0 & \nod \\ 
J0813+2545 (1) & Ku & 08:13:31.27+025.45.3.1 & 0.0487 $\pm$ 0.0031 & 0.0623 $\pm$ 0.0064 & $\leq$151.1$\pm$9.7 & $\leq$148.4$\pm$9.4 & 17.0$\pm$143.0 & \(-1.131\pm0.066\)  \\ 
J0813+2545 (2) & Ku & 08:13:31.33+025.45.2.9 & 0.0215 $\pm$ 0.0030 & 0.0229 $\pm$ 0.0054 & $\leq$143.0$\pm$21.0 & $\leq$131.0$\pm$17.0 & 64.0$\pm$61.0 & \(-1.405\pm0.089\) \\ 
J0813+2545 (3) & Ku & 08:13:31.28+025.45.2.9 & 0.0216 $\pm$ 0.0031 & 0.0338 $\pm$ 0.0074 & $\leq$202.0$\pm$34.0 & $\leq$136.0$\pm$17.0 & 133.0$\pm$12.0 & \(-0.884\pm0.082\) \\  \vspace{0.20cm} 
J0813+2545 (4) & Ku & 08:13:31.32+025.45.3.7 & 0.0056 $\pm$ 0.0030 & 0.0200 $\pm$ 0.0130 & 296.0$\pm$199.0 & 121.0$\pm$127.0 & 114.0$\pm$36.0 & \(-1.131\pm0.297\) \\  
J0911+0550 (2) & C & 09:11:27.63+005.50.54.3 & 0.0117 $\pm$ 0.0053 & 0.0270 $\pm$ 0.0170 & 608.0$\pm$446.0 & 175.0$\pm$245.0 & 93.0$\pm$27.0 & \(-1.328\pm0.291\) \\ 
J0911+0550 (3) & C & 09:11:27.61+005.50.54.9 & 0.0241 $\pm$ 0.0048 & 0.0350 $\pm$ 0.0120 & $\leq$637.0$\pm$184.0 & $\leq$262.0$\pm$36.0 & 117.5$\pm$5.6 & \(-1\pm0.199\) \\ 
J0911+0550 (4) & C & 09:11:27.57+005.50.54.4 & 0.0211 $\pm$ 0.0046 & 0.0210 $\pm$ 0.0088 & $\leq$503.0$\pm$163.0 & $\leq$229.0$\pm$34.0 & 122.5$\pm$6.9 & \(-0.277\pm0.215\) \\ 
J0911+0550 (2) & Ku & 09:11:27.63+005.50.54.2 & 0.0136 $\pm$ 0.0024 & 0.0080 $\pm$ 0.0031 & $\leq$152.0$\pm$34.0 & $\leq$101.0$\pm$14.0 & 143.0$\pm$13.0 & \nod \\ 
J0911+0550 (3) & Ku & 09:11:27.62+005.50.54.8 & 0.0122 $\pm$ 0.0030 & 0.0140 $\pm$ 0.0057 & $\leq$189.0$\pm$51.0 & $\leq$156.0$\pm$35.0 & 133.0$\pm$47.0 & \nod \\  \vspace{0.20cm} 
J0911+0550 (4) & Ku & 09:11:27.57+005.50.54.4 & 0.0113 $\pm$ 0.0026 & 0.0163 $\pm$ 0.0064 & $\leq$331.0$\pm$118.0 & $\leq$113.0$\pm$16.0 & 95.8$\pm$4.5 & \nod \\ 
J0940+3346 (2) & C & 09:40:07.42+033.46.9.5 & 0.0590 $\pm$ 0.0040 & 0.0597 $\pm$ 0.0071 & $\leq$421.0$\pm$34.0 & $\leq$305.0$\pm$18.0 & 99.0$\pm$7.2 & \(-1.285\pm0.113\) \\  \vspace{0.20cm} 
J0940+3346 (2) & Ku & 09:40:07.42+033.46.9.5 & 0.0187 $\pm$ 0.0029 & 0.0184 $\pm$ 0.0053 & $\leq$203.0$\pm$43.0 & $\leq$107.0$\pm$12.0 & 104.3$\pm$6.7 & \nod \\ 
J1010+5705 (1) & C & 10:10:51.14+057.05.30.8 & 0.1100 $\pm$ 0.0044 & 0.1189 $\pm$ 0.0084 & 175.0$\pm$89.0 & 14.0$\pm$65.0 & 77.0$\pm$42.0 & \(-0.693\pm0.057\) \\  \vspace{0.20cm} 
J1010+5705 (1) & Ku & 10:10:51.15+057.05.30.8 & 0.0359 $\pm$ 0.0049 & 0.0630 $\pm$ 0.0130 & 147.0$\pm$56.0 & 70.0$\pm$39.0 & 28.0$\pm$42.0 & \nod \\ 
J1118+0745 (1) & C & 11:18:16.95+007.45.58.0 & 0.0404 $\pm$ 0.0049 & 0.0311 $\pm$ 0.0073 & $\leq$295.0$\pm$40.0 & $\leq$245.0$\pm$27.0 & 91.0$\pm$24.0 & \(0.227\pm0.105\) \\ 
J1118+0745 (2) & C & 11:18:16.96+007.45.58.5 & 0.0331 $\pm$ 0.0055 & 0.0410 $\pm$ 0.0110 & $\leq$405.0$\pm$78.0 & $\leq$287.0$\pm$41.0 & 85.0$\pm$16.0 & \(0.044\pm0.094\) \\ 
J1118+0745 (3) & C & 11:18:16.98+007.45.58.0 & 0.0234 $\pm$ 0.0045 & 0.0140 $\pm$ 0.0056 & $\leq$253.0$\pm$53.0 & $\leq$222.0$\pm$39.0 & 64.0$\pm$53.0 & \nod \\ 
J1118+0745 (1) & Ku & 11:18:16.95+007.45.58.0 & 0.0288 $\pm$ 0.0044 & 0.0383 $\pm$ 0.0094 & 98.0$\pm$45.0 & 29.0$\pm$61.0 & 65.0$\pm$37.0 & \nod \\  \vspace{0.20cm} 
J1118+0745 (2) & Ku & 11:18:16.96+007.45.58.5 & 0.0418 $\pm$ 0.0040 & 0.0427 $\pm$ 0.0074 & $\leq$160.0$\pm$19.9 & $\leq$96.7$\pm$7.3 & 42.8$\pm$5.8 &  \nod \\  \vspace{0.20cm} 
J2246+0344 (1) & C & 22:46:11.24+003.44.17.0 & 0.0279 $\pm$ 0.0044 & 0.0202 $\pm$ 0.0063 & $\leq$277.0$\pm$48.0 & $\leq$237.0$\pm$34.0 & 121.0$\pm$37.0 & \nod \\ 
J2333+2854 (1) & C & 23:33:48.79+028.54.13.0 & 0.0431 $\pm$ 0.0037 & 0.0426 $\pm$ 0.0082 & $\leq$947.0$\pm$158.0 & $\leq$251.0$\pm$11.0 & 64.9$\pm$1.0 & \nod \\ 
J2333+2854 (2) & C & 23:33:48.79+028.54.12.1 & 0.0365 $\pm$ 0.0040 & 0.0413 $\pm$ 0.0091 & $\leq$957.0$\pm$186.0 & $\leq$285.0$\pm$18.0 & 55.0$\pm$1.6 & \nod \\    
\enddata
\tablecomments{ 
(1) Target designation for a pair corresponding to Table \ref{tab:VLA_obs} with primary, secondary, tertiary and quaternary detections labeled as 1, 2, 3 and 4, respectively;
(2) VLA band, where C-band is $\nu\sim6$ GHz and Ku-band is $\nu\sim15$ GHz;
(3) J2000 coordinate of the identified radio source core;
(4) Peak flux density at the central frequency of the band within the source extent;
(5) Integrated flux density at the central frequency of the band within the source extent;
(6$-$8) Best-fit beam-deconvolved source sizes (FWHMs in marcsec) along the major and minor axes and the position angle of the major axis (degrees East of North). In cases of unresolved point sources (denoted with $\leq$), the values are for the un-deconvolved regions. (9) Two-point radio spectral index for cases where the source is detected in both bands. All errors quoted represent 1$\sigma$ uncertainties which are derived via the correlated noise prescription of \citet{Condon1997}. The error of the photometry does not include the 3\% uncertainty in the VLA flux density scale \citep{Perley2013}.}
\end{deluxetable*}

\begin{figure*}[ht!]
     \centering
         \includegraphics[width=7.1in]{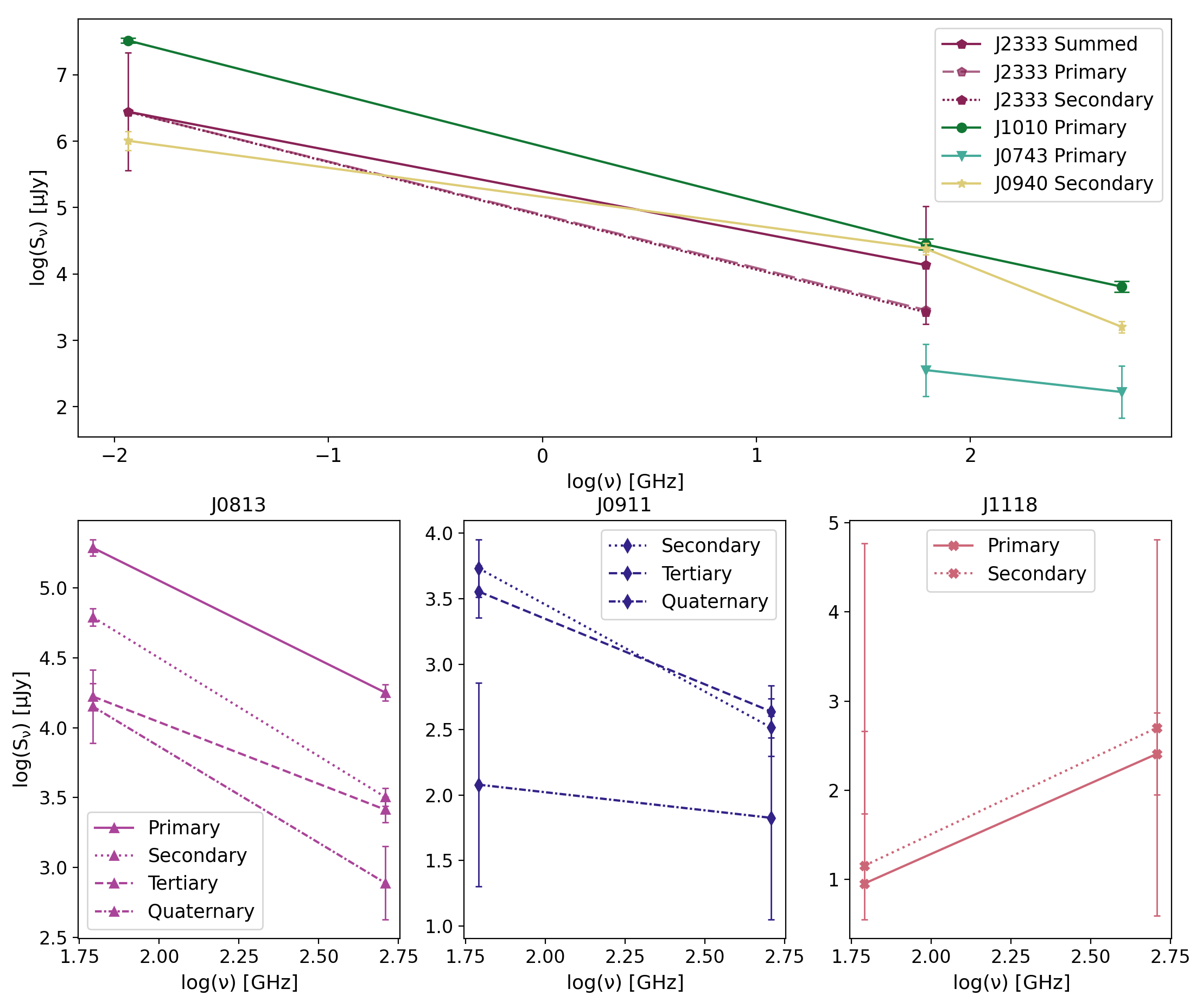}
        \caption{Radio spectral energy distributions of the targets with LOFAR, and VLA C and Ku-band detections. On the top graph, we report the radio spectral energy distributions of J0743+2209 primary, J0940+3346 secondary, J1010+5705 primary, and J2333+2854 detections. J0940+3346 and J1010+5705 are extended to accommodate detections at 144 MHz by LOFAR. The spectral index of J2333+2854 is only calculated between C-band and LOFAR detections. Because J2333+2854 shows detections of the primary and the secondary in C-band, we plot 3 separate cases, corresponding to if J2333+2854 LOFAR flux belongs only to the primary, only to the secondary, or is a blend of both. On the bottom three graphs, the spectral indices of the quadruply lensed detections in J0813+2545, J0911+0550, and J1118+0745 are separately shown. Due to J1118+0745 detections being faint, their spectral indices are poorly constrained and have large uncertainties. We conclude that the majority of the targets have steep spectra, with the exception of J0743+2209 primary, J0911+0550 quaternary and J1118+0745 detections, which have flat spectral indices. It is unusual to see the positive, upturned slope for J1118+0745 primary and secondary.}
     \label{fig:spectral_indices}
\end{figure*}
 

\section{Discussion}\label{sec:discussion}
In this section, we derive various properties based on the observation results. We calculate spectral index, radio loudness, radio luminosity and star formation rate for appropriate targets using the C and Ku-band integrated fluxes, Gaia \textit{G}-band magnitudes and SDSS spectra. We also utilize LOw-Frequency ARray (LOFAR), Hubble Space Telescope (\textit{HST}) and James Webb Space Telescope (\textit{JWST}) detections for select targets as additional lines of evidence. Using the time-domain coadded images of the Wide-field Infrared Survey Explorer observations, we confirm that all dual AGN candidates have at least one infrared AGN \citep{Jarrett2011, Wright2019, Meisner2024}.

\subsection{Radio Emission in AGN}\label{sec:radio}
The flux density of radio sources is expected to follow a power law spectrum over observed frequency:
\begin{equation}\label{index1}
S_{\nu} \propto \nu^{\alpha},
\end{equation}
where \(S_{\nu}\) is the flux density at frequency \(\nu\), and \(\alpha\) is the spectral index. Steep AGN spectrum (\(\alpha < -0.7\)) usually corresponds to synchrotron emission from optically thin regions of the jet \citep{Blandford1979}. Meanwhile, flat or close-to-flat (\(\alpha \sim 0\)) inverted spectra is associated with optically thick jet launching regions emitting synchrotron radiation \citep{Blandford1979}. Therefore, a flat spectrum may correspond to the AGN core, where the jet base is. 

As calculating spectral index requires detections in both bands, we only have this value for J0743+2209 primary, all J0813+2545 detections, J0911+0550 secondary, tertiary and quaternary, J0940+3346 secondary, J1010+5705 primary, and J1118+0745 primary and secondary. We plot the spectral energy distributions calculated between 144 MHz, 6 GHz and 15 GHz when possible on Figure \ref{fig:spectral_indices}.

J0813+2545, J0940+3346, J1010+5705 and J2333+2854 have LOFAR Two-metre Sky Survey detections. The images were observed with the LOFAR High Band Antenna and are centered at 144 MHz. Their resolution is 6\arcsec \ with rms sensitivity of 83 \(\mu \mathrm{Jy \ beam^{-1}}\) \citep{LOFAR2013}. We do not use the LOFAR data for J0813+2545 as it is a quadruply lensed AGN \citep{Liu2026}.

Due to the resolution of the LOFAR images and the low separation of the targets, the LOFAR images do not resolve the primary and the secondary. It is possible to estimate the expected fluxes of the primary and the secondary based on spectral energy distributions between VLA C and Ku-bands. Because J0940+3346 and J1010+5705 only show a single detection at 6 and 15 GHz, we assume they also have single detections at 144 MHz. The LOFAR flux of J0940+3346 and J1010+5705 are \(0.6\pm 0.1 \ \mathrm{mJy}\) and \(2.3\pm 0.2\ \mathrm{mJy}\), respectively. The total LOFAR flux of J2333+2854 is \(0.9\pm 0.2\ \mathrm{mJy}\). For this target with 2 detections at 6 GHz and no detection at 15 GHz, we consider three cases: 
\begin{itemize}
    \item LOFAR flux belongs entirely to the primary.
    \item LOFAR flux belongs entirely to the secondary.
    \item LOFAR flux has contribution from both the primary and the secondary.
\end{itemize}
The first two cases are strict upper or lower limits, and we use them to constrain the error in the spectral energy distribution of J2333+2854. A precipitous drop in radio flux for either target at both 144 MHz and 15 GHz would yield an unlikely highly peaked radio spectral energy distribution. 

\subsection{Gravitational Lensing}\label{sec:lensing}
Due to the astrometric accuracy of Gaia EDR3 at 0.05 \(\mathrm{mas}\), all targets have precise coordinates and Gaia \textit{G}-band magnitudes \citep{Kozhurina-Platais2021}. Gaia has a resolution of 0\farcs2 and is capable of making accurate measurements of proper motions, which is crucial for identifying closely separated targets. Moreover, the Gaia bandpass ranges from 330 nm to 1050 nm; thus, it might be enclosing prominent quasar emission lines in the observed frame. 

A potential contaminant to dual AGN samples are single AGNs that are gravitationally lensed by intervening structures. For a gravitationally lensed object, the spectra of the primary and the secondary would be the same, besides the spectrum of the primary getting scaled by a magnification factor. We compare the ratios between the fluxes of the primary and the secondary in VLA C and Ku-bands, and Gaia \textit{G}-band to test for gravitational lensing. For lensed targets, we expect the ratios to be significantly similar. We calculate the flux ratios \(F_{Gaia}\) in Gaia \textit{G}-band magnitudes between two targets \(m_1, m_2\), following \(F_{Gaia} = 10^{-0.4\left(m_1 - m_2\right)}\).

We also use spectral indices for the primary and secondary targets as another way of qualitatively testing for gravitational lensing, following Equation \ref{index1}. Significant difference between the spectral indices calculated based on the spectra of the primary and the secondary is another indication that both targets in a pair are unique.

In C-band, two components of the J0813+2545 are blended (labeled VLA 1 and VLA 4 in Figure \ref{fig:Lensed}). Since they are resolved in Ku-band, we use their flux density ratios in Ku-band to separate the blended flux density in C-band into its separate components. We assume the flux ratio should be the same in both bands due to knowing a priori that this is a lensed target \citep{Liu2026}. We obtain \(176\pm50 \ \mathrm{\mu Jy}\) for VLA 1, and \(56\pm 37 \ \mathrm{\mu Jy}\) for VLA 4. Using the calculated flux densities, we obtain spectral indices for all lens components of J0813+2545.

\begin{figure}[h!]
    \centering
        \includegraphics[width = 0.48\textwidth]{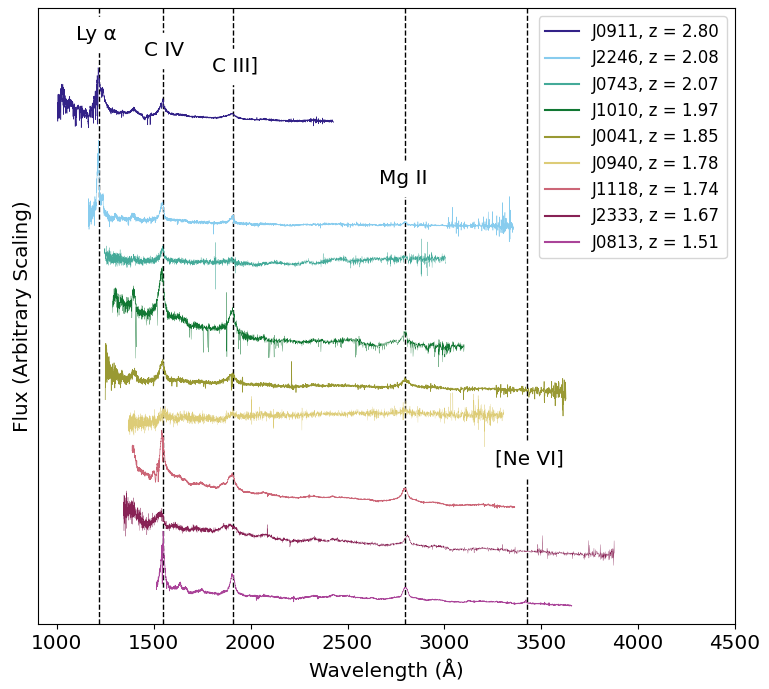}
        \caption{Rest-frame SDSS spectra of all targets, ordered in increasing redshift with arbitrary scaling on the y-axis. Our targets come from a wide range of Cosmic Noon redshifts. The dashed lines correspond to prominent emission lines. Some targets have asymmetrical blueshifted broad emission lines,  such as J2333+2854 on C IV, which  can indicate AGN winds or a superimposed dual AGN spectrum.}
    \label{fig:SDSSspectra}
\end{figure}

\begin{deluxetable}{lccccr}
\tablewidth{0.4\textwidth}
\tabletypesize{\scriptsize}
\tablecaption{Gaia and SDSS Observations
\label{tab:Additional}}
\tablehead{ 
\colhead{Target} & \colhead{SDSS \textit{i}-band} & \colhead{Gaia \textit{G}-band (1)} & \colhead{\textit{G}-band (2)} & \colhead{\(F_{Gaia}\)} \\
\colhead{J200} &\colhead{($\mu$Jy)} & \colhead{(mag)} & \colhead{(mag)} & \colhead{}\\
\colhead{(1)} & \colhead{(2)} & \colhead{(3)} & \colhead{(4)} & \colhead{(5)} }
\startdata 
J0041-0203 & 125.283 & 19.184 & 20.444 & 3.189 \\ 
J0743+2209 & 111.090 & 19.700 & 19.724 & 1.023 \\ 
J0813+2545 & 2755.138 & 16.341 & 17.940 & 4.360 \\ 
J0911+0550 & 264.480 & 19.508 & 18.873 & 0.557 \\ 
J0940+3346 & 127.974 & 19.713 & 19.593 & 0.895 \\ 
J1010+5705 & 617.907 & 17.348 & 19.292 & 5.993 \\ 
J1118+0745 & 1403.710 & 17.215 & 17.089 & 0.891 \\ 
J2246+0344 & 129.111 & 19.222 & 20.295 & 2.685 \\
J2333+2854 & 207.708 & 18.191 & 20.301 & 6.985 \\ 
\enddata
\tablecomments{ 
(1) Target Designation;
(2) SDSS \textit{i}-band Petrosian flux in \(\mathrm{\mu Jy}\);
(3) Gaia \textit{G}-band magnitude of the primary detection;
(4) Gaia \textit{G}-band magnitude of the secondary detection;
(5) Gaia \textit{G}-band flux ratio between the primary and secondary detections.}
\end{deluxetable}

\subsection{Radio Loudness}\label{sec:RL}
Radio loudness is a metric commonly used to assess how strongly a source emits in radio wavelengths relative to optical wavelengths. For both Type I and II AGN in radiative mode, it is possible to have radio quiet and radio loud AGNs with strong optical emission \citep{Heckman2014}. Furthermore, recent models suggest that radio jets can be prominent in both the low accretion and super-Eddington regimes, corresponding to jet mode and radiative mode AGN, respectively \citep{Heckman2014, Husko2026}. Determining whether an AGN is radio loud or radio quiet informs us of which radiative processes dominate \citep{Panessa2019}, which is useful for high redshift sources where radio morphology does not clearly expose jet structure.

We calculate radio loudness for each target using \textit{i}-band Petrosian flux from SDSS data, which encapsulates both the primary and the secondary targets as SDSS does not spatially resolve both Gaia detections. In Figure \ref{fig:SDSSspectra}, we show the rest-frame SDSS spectrum of each target. The PCA analysis done by \citet{Shen2023} did not reveal any obvious dual quasar spectra. Thus, if any of the targets are indeed dual optical quasars, the lack of obvious double-peaked emission lines suggests that the systematic redshifts of each component AGN must be very similar. The Gaia \textit{G}-band magnitudes and their ratios, as well as SDSS \textit{i}-band Petrosian fluxes for the targets are given in Table \ref{tab:Additional}.

We adopt the following convention from \citet{Jiang2007} for calculating radio loudness:
\begin{equation}\label{RL}
R = \frac{S_{\mathrm{6 \ cm}}}{S_{2500 \textup{~\AA}}},
\end{equation} where \(R\) is the radio loudness, \(S_{\mathrm{6 \ cm}}\) is flux density at 6 cm, which corresponds to 5 GHz, and \(S_{2500 \textup{~\AA} }\) is flux density at 2500 \AA \ \citep{Jiang2007}. \(R > 1\) points to a radio loud source, while \(R < 1\) points to a radio quiet source.

We obtain \(S_{\mathrm{6 \ cm}}\) by extrapolating from the measured flux densities at 6 GHz, corresponding to C-band flux densities, for targets with spectral indices. We assume \(\alpha = -0.7\) \citep{Condon2002}, the canonical AGN spectral index, for J2246+0344 and J2333+2854 for which we could not calculate spectral index due to lack of detections in Ku-band. We extract \(S_{2500 \textup{~\AA} }\) from the target spectra at rest wavelength closest to 2500 \AA.

J0911+0550 has the highest redshift among all targets, with \(z = 2.8\). This redshift is too high for the de-redshifted spectrum to extend to 2500 \AA, it only extends to 2427 \AA. We fit a second-order polynomial to the last 40 bins of the spectrum using \texttt{scipy}, and construct a function that extrapolates the spectrum to higher wavelengths. The obtained flux density at 2500 \AA \ is 11\% different than the preceding 40 bins. Based on Figure \ref{fig:SDSSspectra}, we are confident this extracted flux is not affected by prominent emission lines.

Since \(S_{2500 \textup{~\AA} }\) extracted from SDSS spectra encloses both the primary and secondary detections in optical band, we separate the SDSS flux based on Gaia flux ratios of individual targets when possible. Gaia \textit{G}-band (ranging from 330 nm to 1050 nm) coincides with the central wavelength of SDSS \textit{i}-band 748 nm, appropriate to do this rough empirical rescaling \citep{Fukugita1996}. For J0743+2209, J0940+3346, J1010+5705, J2246+0344 and J2333+2854, we report the radio loudness calculated with split SDSS fluxes. For J0813+2545, J0911+0550 and J1118+0745, we report the radio loudness calculated with the unsplit SDSS flux and total radio flux of the lens components as they are known quad lenses. We expect J0911+0550 and J1118+0745 to be radio quiet as not all lens components seen in optical and infrared wavelengths are detected in radio band. We cannot calculate radio loudness for J0041-0203 due to lack of detections in both radio bands. We report the radio luminosity and radio loudness for appropriate targets in Table \ref{tab:Derived}.

Across all detections for which we could calculate radio loudness, we find that the majority are radio quiet: J0813+2545, J0911+0550, J1010+5705, J1118+0745, J2246+0344 and J2333+2854 primary. It is possible that the radio quiet detections which are verified AGNs, J0813+2545, J1010+5705 primary and J1118+0745 might have radio emission from the corona. J0743+2209, J0940+3346 and J2333+2854 secondary are radio loud. Even though the spatial resolution does not show jet structure, we infer these detections belong to AGN jets. Furthermore, launching new radio jets depends on contribution from fading jets, especially in merger scenarios \citep{An2026}. This introduces time-domain radio loudness where radio loudness is dependent on core emission from new jets and the previously launched jets \citep{An2026}. Therefore, radio-quietness and flat spectral index observed in this sample suggest that radio emission must originate from recent activity, such as a merger.

\subsection{Star Formation Rate}\label{sec:SFR}
Cosmic Noon is a period of the Universe with high star formation rates (SFRs) \citep{Madau2014}. High SFRs increase the rate of supernovae, which create synchrotron emission by acceleration of cosmic rays through the resulting shock fronts. This leads to star-forming regions and starburst galaxies having steep radio spectra, a signature which can be confused with non-thermal synchrotron emission attributed to AGN jet activity when jets cannot be resolved \citep{Panessa2019}. Hence, we calculate SFRs for targets with steep spectra in order to see if it is realistic for the target to be explained better by an AGN or a star-forming region with its observed radio flux density. In most cases, true AGNs have flux densities much higher than star-forming regions, which lead to SFRs beyond what is expected at Cosmic Noon (\(\sim\)100 \(\mathrm{M_\odot \ yr^{-1}}\)) \citep{Forster2020}.

The spectral index associated with star-forming regions is \(-0.8\), which we assume for all targets for the subsequent fiducial analysis. Following the method of \citet{Zakamska2016}, we first calculate the flux density at \(1.4 \ \mathrm{GHz}\) by extrapolating the Ku-band flux density \(S_{15 \mathrm {GHz}}\) to \(S_{1.4 \mathrm{GHz}}\). We use Ku-band as it is more compact and less likely to be due to extended star-forming regions. We then obtain the luminosity at 1.4 \(\mathrm{GHz}\) to calculate SFR based on \citet{Bell2003}.

As J2246+0344 primary and both J2333+2854 targets do not have detections in Ku-band, we instead use their C-band flux densities to extrapolate to flux densities at 1.4 \(\mathrm{GHz}\), using the same method from \citet{Zakamska2016}. The calculated SFR values for the dual AGN candidates range from 915 \(\mathrm{M_\odot \ yr^{-1}}\) to 5220 \(\mathrm{M_\odot \ yr^{-1}}\), which is well above the predicted SFR for star-forming regions at Cosmic Noon \citep{Daddi2005, Madau2014}. Thus, we conclude that none of the radio sources are due to star formation, except for the lensing galaxy in the J0911+0550 quad lens system. 

\begin{deluxetable}{lccrrrrrr}
\tablewidth{\textwidth}
\tabletypesize{\scriptsize}
\tablecaption{Derived Properties of Radio Detections
\label{tab:Derived}}
\tablehead{ 
\colhead{Target} & \colhead{Detection}&  \colhead{\(L_{6\mathrm{GHz}}\)}&\colhead{\(L_{15\mathrm{GHz}}\)} & \colhead{\(R_{split}\)} & \colhead{\(R_{summed}\)} \\
\colhead{J2000}  & \colhead{}& \colhead{(erg/s/Hz)}& \colhead{(erg/s/Hz)} & \colhead{}  & \colhead{} \\
\colhead{(1)} & \colhead{(2)} & \colhead{(3)} & \colhead{(4)} & \colhead{(5)} & \colhead{(6)}
}
\startdata 
J0743+2209 & 1 & \(4.016\times10^{30}\) & \(2.886\times10^{30}\) & 1.079 & \nod \\ 
J0813+2545 & All & \(6.384\times10^{31}\) & \(2.269\times10^{31}\) & \nod & 0.174 \\ 
J0911+0550 & All & \(8.394\times10^{31}\) & \(3.873\times10^{31}\) & \nod & 0.274 \\
J0940+3346 & 2 & \(1.739\times10^{31}\) & \(5.360\times10^{30}\) & 8.398 & \nod \\ 
J1010+5705 & 1 & \(2.372\times10^{31}\) & \(1.257\times10^{31}\) & 0.959 & \nod \\
J1118+0745 & All & \(4.288\times10^{30}\) & \(4.817\times10^{30}\) & \nod & 0.049 \\
J2246+0344 & 1 & \(4.602\times10^{30}\) & \nod & 0.244 & \nod \\
J2333+2854 & 1 & \(5.918\times10^{30}\) & \nod & 0.266 & \nod \\
J2333+2854 & 2 & \(5.738\times10^{30}\) & \nod & 1.804 & \nod
\enddata
\tablecomments{ 
(1) Target designation for a pair corresponding to Table \ref{tab:VLA_obs};
(2) Radio detection for the target designation: 1 - primary, 2 - secondary, 3 - tertiary, 4 - quaternary;
(3) Radio luminosity of the detection at \(6 \mathrm{GHz}\), corresponding to VLA C-band;
(4) Radio luminosity of the detection at \(15 \mathrm{GHz}\), corresponding to VLA Ku-band;
(5) Radio loudness of the radio detection calculated using split SDSS \textit{i}-band flux; (6) Radio loudness of the radio detection calculated without splitting SDSS \textit{i}-band flux and summing radio fluxes corresponding to different lens components.
}
\end{deluxetable}

\subsection{Discussion of Individual Targets}\label{sec:individual}
\subsubsection{J0041-0203}\label{sec:J0041}
J0041-0203 showed no detection in either band. Its primary is a confirmed AGN, based on SDSS and WISE observations. It is possible that J0041-0203 secondary is too faint in radio band to be significantly detected, or corresponds to a galaxy or star.

\subsubsection{J0743+2209}\label{sec:J0743} 
\citet{Shen2023} tentatively identified J0743+2209 as a quasar-star superposition based on PCA. The spectrum of J0743+2209 showed absorption lines consistent with that of an M3-type star combined with typical AGN spectrum. 

The primary has a flat spectral index. This indicates synchrotron self-absorption from AGN core region \citep{Panessa2019}. Since the target has flat spectral index and is not spatially resolved, we rule out star forming region as a possible explanation for the radio emission.

In both C and Ku-bands, emission from only the primary target was detected. This strengthens the Gaia classification for the primary as an AGN, and the argument by PCA for the secondary as a star.

\subsubsection{J0813+2545}\label{sec:J0813}
J0813+2545 was identified as a quadruply lensed AGN based on preliminary \textit{JWST} imaging after its inclusion in the sample for this project \citep{Liu2026}. The color composite image for J0813+2545 using three \textit{JWST} filters is shown on Figure \ref{fig:Lensed}. 

The quad lens is equally confirmed by the flux ratios between the primary and the secondary in each band and in Gaia detections, as well as the difference between the spectral indices of the different components. The flux ratio between the primary and secondary is \(2.8 \pm 0.4\) in C-band, and \(2.7 \pm 0.7\) in Ku-band, corresponding to a difference of 44\% between C-band and Gaia observation, and 46\% in between Ku-band and Gaia observation. Furthermore, we calculate the spectral index of the quaternary in C-band by extrapolating from the Ku-band values and find that the spectral indices are less than 50\% different between all components. Thus, we confirm that the flux densities scale the same across rest-frame optical and radio bands, consistent with gravitational lensing.

All spectral indices calculated for J0813+2545 fall below \(-0.5\), corresponding to steep spectral indices, and indicate non-thermal synchrotron emission. To confirm J0813+2545 is an AGN and not a starburst galaxy, we calculate the SFR required with its observed flux density for the primary to be 2770 \(\mathrm{M_\odot \ yr^{-1}}\), which is higher than the typical SFR for a star-forming galaxy at Cosmic Noon. Since we do not know the magnification factor of the lensing, the calculated SFR serves as an upper limit. Moreover, a high magnification factor would still yield an intrinsic SFR that is above expectations for this epoch. Consequently, we confirm that J0813+2545 is a single, gravitationally lensed AGN.

\subsubsection{J0911+0550}\label{sec:J0911}
\begin{figure}
    \centering
         \includegraphics[width = 0.45\textwidth]{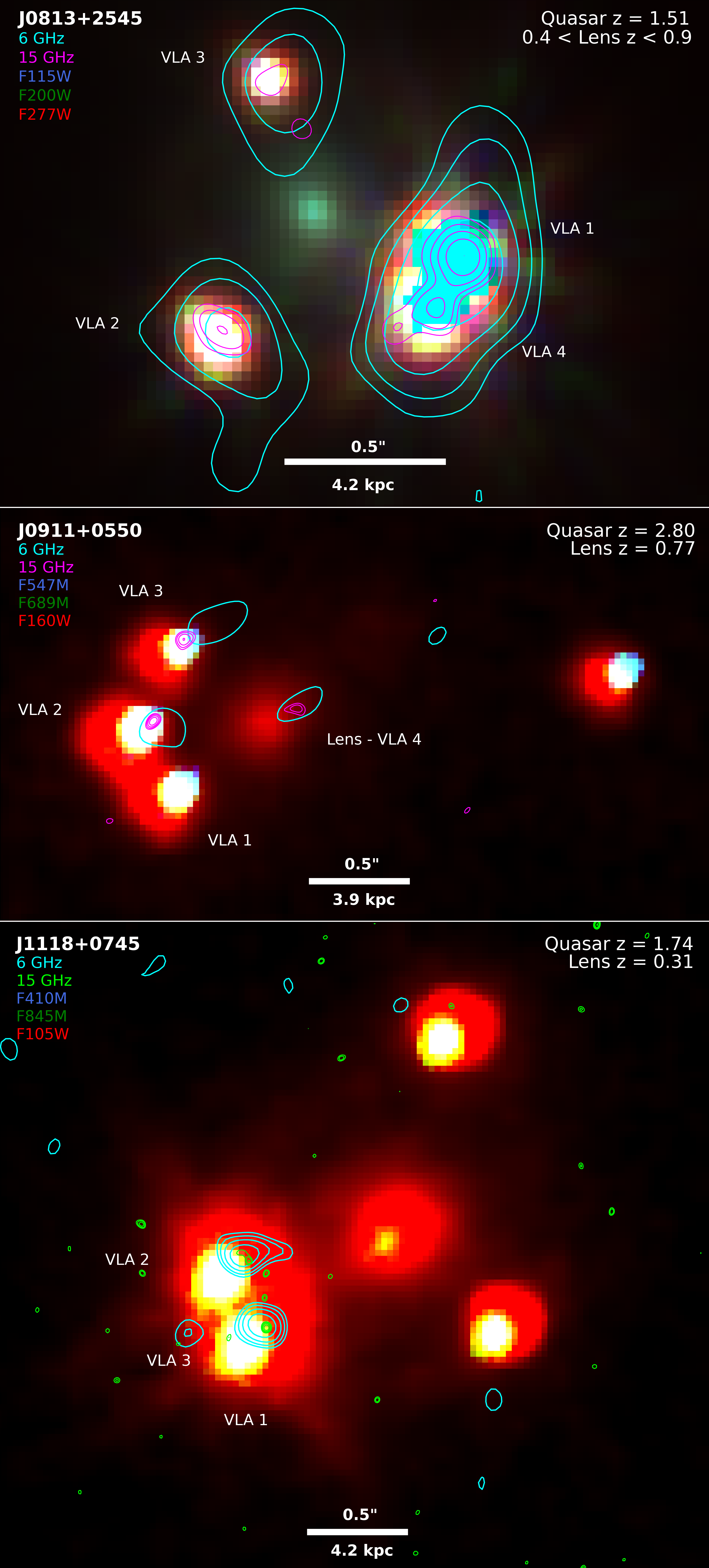}
        \caption{Color composite images of the three quadruply lensed targets: J0813+2545, J0911+0550 and J1118+0745. All images are oriented such that North is up and East is left. Each image has the components which were detected by either VLA band labeled, referring to VLA primary, secondary, tertiary and quaternary components. For J0813+2545, we use \textit{JWST} images from 3 filters. The blue and green filters show oversaturation in pixels corresponding to the primary and quaternary components. For J0911+0550 and J1118+0745, we use \textit{HST} images from 3 filters. For all three images, we show contours from VLA C (6 GHz) and Ku-bands (15 GHz). Because VLA has stronger absolute astrometry compared to \textit{HST}, there are linear offsets between the image components and the VLA contours. For J0911+0550 and J1118+0745, the radio flux anomaly is readily apparent with bright optical sources having no radio counterparts.}
     \label{fig:Lensed}
\end{figure}

J0911+0550 is identified to be a quadruply lensed AGN by \citet{Burud1998}. Based on the gravitational lens model, the VLA detections identified as secondary and tertiary are two of the lens components. The quaternary VLA detection is consistent with the coordinates of the lensing galaxy with \(z = 0.77\) \citep{Kneib2000}. The fourth component of the quad lens is approximately 1\farcs65 away from the other three and the lens. There is no significant detection in C or Ku-band at the coordinates of the fourth lens component. Moreover, the primary, as identified by Gaia observations, is not significantly detected in either VLA band (\(2.5\sigma\) in C-band, no detection in Ku-band). Gaia \textit{G}-band magnitudes of the two lens components not detected in either VLA band are 19.7 for the tertiary and 19.9 for the fourth lens component. In Gaia \textit{G}-band, the primary-to-secondary flux ratio is 0.6, the primary-to-tertiary ratio is 1.2, and the primary-to-fourth-lens component ratio is 1.4. Adopting the marginal $2.5\sigma$ C-band primary, the corresponding radio flux ratios are 0.7 (secondary) and 0.6 (tertiary). Rerunning \texttt{tclean} with different parameters also yielded no significant detections for these targets, which proves that it is not due to instrumental effects. In Figure \ref{fig:Lensed}, we plot the color composite image for J0911+0550 using images from three \textit{HST} filters with VLA contours overlaid to show the flux inconsistency across different wavelengths.

Only the spectral indices of the secondary, the tertiary and the quaternary detections were possible to calculate. The tertiary and the secondary show steep spectra, indicating non-thermal synchrotron emission, which is consistent for a single AGN getting gravitationally lensed. The spectral index difference is 28\% between the tertiary and the secondary.

We take the VLA quaternary to be the lensing galaxy, which has an SFR of 147 \(\mathrm{M_\odot \ yr^{-1}}\) at \(z = 0.77\) and flat spectral index. The lens has been identified to be a massive galaxy cluster by \citet{Kneib2000}, which is consistent with the chromatic effects caused by microlensing of J0911+0550 even though it has a flat spectral index in radio band. 

The lack of significant detections for the VLA primary and the fourth lens component in both bands is against what is expected of achromatic gravitational lensing. We consider it to be a case of radio flux anomaly. The chromatic effects in optical, infrared and radio wavelengths for J0911+0550 have been documented before \citep{Yonehara2008, Jackson2015, Perera2023, Fores2024}. Studies identify the possible causes to be either microlensing or dust, with the former being more likely in the case of J0911+0550 based on lens modeling \citep{Yonehara2008, Jackson2015}. The VLA observations support the identification of J0911+0550 as a quadruply lensed single AGN and show more evidence towards the previously suggested radio flux anomaly.

\subsubsection{J0940+3346}\label{sec:J0940}
J0940+3346 shows only the detection of the secondary in both bands. The spectral index of the secondary is \(-1.3\), a strong steep spectrum pointing to non-thermal synchrotron emission. In order to rule out the possibility of a starburst galaxy, we calculate the corresponding SFR and find J0940+3346 secondary must be forming stars at a rate of 1240 \(\mathrm{M_\odot \ yr^{-1}}\) to be producing its observed flux density, which is unlikely. 

As the primary has been identified as an AGN based on Gaia observations and SDSS spectroscopy, we conclude J0940+3346 to be a dual AGN system. It contains one optical quasar, the primary, and one radio AGN, the secondary.

\subsubsection{J1010+5705}\label{sec:J1010}
J1010+5705 shows the detection of only the primary in both bands with a steep spectral index. To understand the mechanism behind the non-thermal synchrotron emission, we calculate its SFR to be 5220 \(\mathrm{M_\odot \ yr^{-1}}\), which is unlikely. Thus, the synchrotron emission inferred from the steep spectrum is attributed to an AGN. 

The SDSS spectrum of J1010+5705 shows many absorption lines at the same redshift; however, we cannot conclude any further results about the J1010+5705 secondary. It could be a radio quiet AGN, a galaxy, or a star with the primary as a confirmed AGN.

\subsubsection{J1118+0745}\label{sec:J1118}
J1118+0745 was identified to be a quadruply lensed AGN by \citet{Impey1998}. It shows two detections in both bands, as well as a third significant detection in C-band. This detection is not observed in Ku-band or by \textit{HST} which implies that it was not lensed. Thus, it must have a lower redshift compared to the lens (\(z = 0.311 \) \citealt{Kundic1997}). Taking this redshift as the upper limit, we calculate its SFR to be 15.4 \(\mathrm{M_\odot \ yr^{-1}}\). Therefore, we consider the possibility that it is a small and faint foreground galaxy.

The flux ratio of the primary to the secondary is \(0.8 \pm 0.3\) in C-band and \(1 \pm 0.3\) in Ku-band. This corresponds to a difference of 16\% between Gaia observation and C-band, and 1\% between Gaia observation and Ku-band. These statistics are consistent with the identification of J1118+0745 as a quad lens. On the other hand, the percent difference of the spectral indices is 135\%. The primary is the brighter object in C-band, while the secondary is the brighter object in Ku-band. Moreover, the third and fourth lens components described in \citet{Impey1998} are not observed in either VLA band. It is confirmed by rerunning \texttt{tclean} with different parameters that the lack of detections and the significant third detection in C-band are not dependent on image reduction effects. In Figure \ref{fig:Lensed}, we plot the color composite image for J1118+0745 using images from three \textit{HST} filters with VLA contours overlaid to show the flux inconsistency across different wavelengths. Like J0911+0550, we observe radio flux anomaly in J1118+0745. The flux inconsistency between radio, optical and X-ray observations have been known from previous studies \citep{Impey1998, Pooley2006, Hartley2021} and microlensing is considered to be the cause of these chromatic effects \citep{Yonehara2008, Tsvetkova2010, Takahashi2014, Fores2024}. Hence, we confirm J1118+0745 as a quadruply lensed single AGN showing radio flux anomaly. 

\subsubsection{J2246+0344}\label{sec:J2246}
J2246+0344 has a \(4.7\sigma\) detection of only the primary in C-band, and no detection of the secondary in either band. This suggests that there is no gravitational lensing as the primary would have been detected in Ku-band, with the given rms levels in Section \ref{sec:method}. The SDSS spectrum of J2246+0344 shows that the C IV broad emission line is blueshifted, which likely indicates outflows from AGN winds. 

Due to lack of detections in Ku-band, it was not possible to calculate flux ratios or spectral indices. Assuming J2246+0344 is a star-forming region, we calculate the SFR to be 915 \(\mathrm{M_\odot \ yr^{-1}}\), which gives an unlikely result. The J2246+0344 primary is a confirmed AGN based on SDSS and VLA observations, with possibly a radio quiet AGN, a star or a galaxy as the secondary.

\subsubsection{J2333+2854}\label{sec:J2333}
J2333+2854 shows two detections in C-band and no detection in Ku-band. Thus, we were not able to compute spectral indices. The flux ratio between the primary and secondary is \(1.0 \pm 0.3\) in C-band.

Using \(\alpha = -0.8\) for the assumed star-forming region spectral index, we calculate the SFRs for J2333+2854 primary and secondary. The primary has a SFR of 1160 \(\mathrm{M_\odot \ yr^{-1}}\) and the secondary has a SFR of 1120 \(\mathrm{M_\odot \ yr^{-1}}\), proving to be too high for a star-forming region at Cosmic Noon.

There is no ``smearing" due to the restoring beam between the primary and the secondary, they are oriented such that they do not blend. \citet{Ji2023} identified J2333+2854 as a lensed quasar based on Gaia and SDSS observations alone. Our VLA C-band imaging reveals two spatially distinct radio detections whose C-band flux ratio differs by $149\%$ from the Gaia optical flux ratio. Under gravitational lensing, flux ratios are expected to be achromatic (modulo microlensing effects); a discrepancy of this magnitude is difficult to reconcile with a simple lensing geometry. We therefore favor the dual AGN interpretation, though we note that the absence of a Ku-band detection prevents us from computing spectral indices and leaves the classification less secure than for J0940+3346. 


\section{Summary \& Conclusion}\label{sec:conclusion}
In this work, we try to identify dual AGNs at Cosmic Noon, with redshifts ranging from 1.5 to 2.8, in order to answer if galaxy mergers cause one or both SMBHs to become active. The sample was chosen based on spectroscopically confirmed SDSS DR16 quasars. Among these quasars, those with a secondary detection within 3\arcsec \ radius circle in Gaia EDR3 were identified for analysis in radio band. Nine candidates out of this sample received VLA imaging in both C and Ku-bands, which we discuss in this paper. All of these targets have their primary detection confirmed to be an AGN, while we try to understand the nature of the secondary targets. Of the nine targets, prior optical or near-infrared imaging had already classified J0813+2545, J0911+0550 and J1118+0745 as quadruply lensed AGNs, and J0743+2209 as a likely AGN-star superposition. The VLA observations confirm these classifications and additionally identify J0940+3346 and J2333+2854 as new dual AGN candidates, while revealing radio flux anomalies in two of the three lensed systems.

\begin{figure}[h!]
    \centering
         \includegraphics[width = 0.45\textwidth]{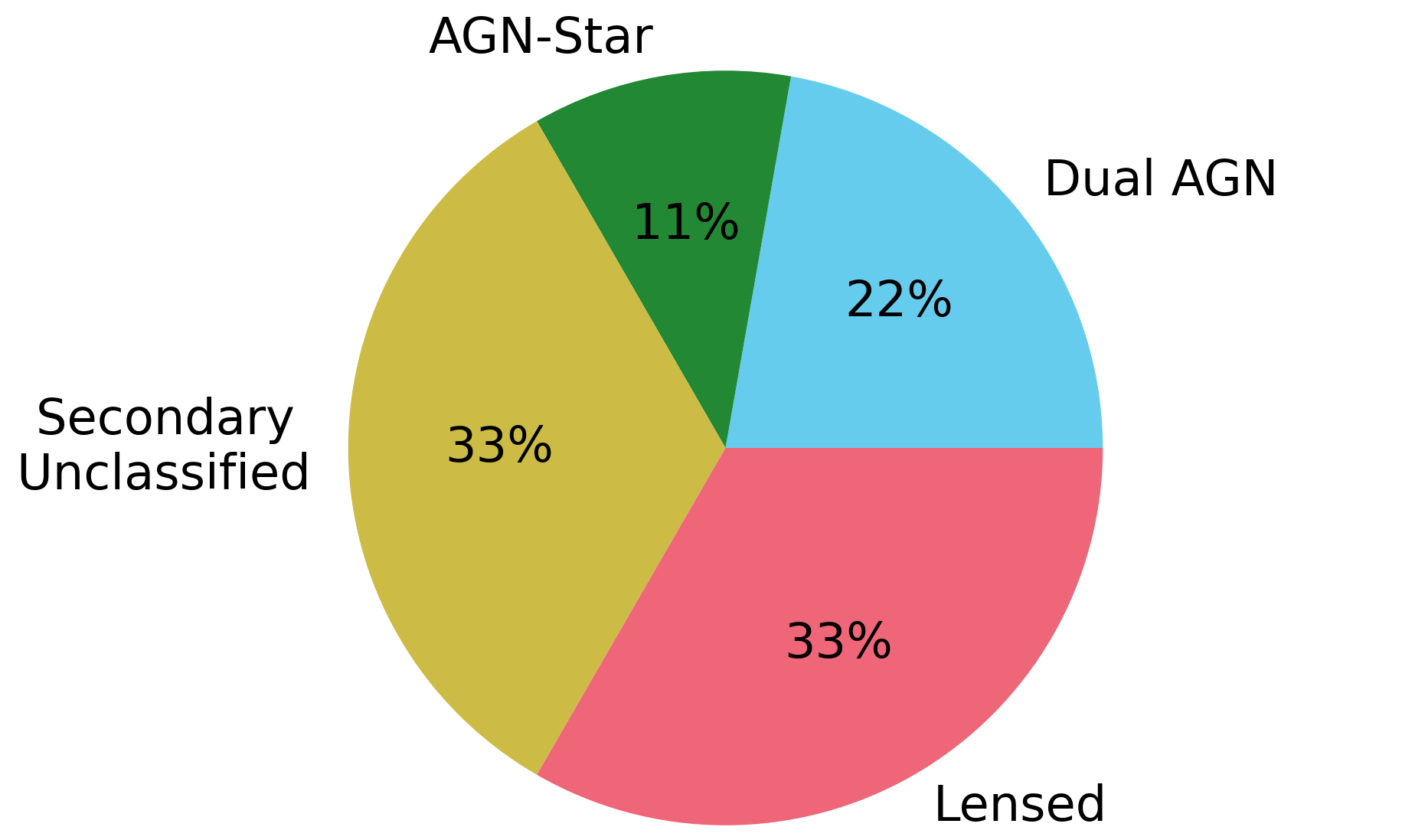}
        \caption{Distribution of final target classifications for the nine VLA-observed candidates. The high lensed fraction (33\%) highlights the importance of multi-wavelength follow-up for Gaia-selected high-$z$ pairs with small separations.}
     \label{fig:pie}
\end{figure}

Therefore, we identify two of the nine targets to be true dual AGNs: J0940+3346 and J2333+2854. We classify one target to be an AGN-star pair, J0743+2209. J0041-0203, J1010+5705 and J2246+0344 have at least one confirmed AGN, the primary Gaia detections, while their secondary detections remain unclassified. We identify three targets, J0813+2545, J0911+0550 and J1118+0745, to be single quadruply lensed AGNs. Among the nine targets, both of the confirmed dual AGNs (J0940+3346 and J2333+2854) show evidence of ongoing or recent mergers, with one member of each pair detected only in radio and not in optical band, consistent with dust-obscured AGN activity triggered by the merger. Given the small sample size and the specific selection criteria favoring previously unstudied, small-separation pairs, these fractions, i.e., roughly 22\% dual AGN and 33\% lensed, should be treated as illustrative rather than statistically robust. While the sample is too small to draw statistically robust conclusions, these results are qualitatively consistent with the prediction that mergers can trigger AGN activity in both nuclei simultaneously \citep{DiMatteo2005, VanWassenhove2012}. Analysis of the remaining 21 candidates currently receiving VLA C-band imaging will be critical for a more quantitative assessment. We illustrate the distribution of results in Figure \ref{fig:pie}. 

Two out of the three quadruply lensed AGNs, J0911+0550 and J1118+0745, show clear indication of radio flux anomaly. Gravitational microlensing is considered to be the cause of the radio flux anomaly in both cases. It is theorized that microlensing is a result of cold dark matter subhaloes in the lensing galaxy halo \citep{Mao1998, Metcalf2001, Dalal2002}. However, recent findings suggest that subhaloes alone cannot explain the discrepancy between magnification ratios across different wavelengths, and that structures in the line-of-sight equally contribute to microlensing effects \citep{Chen2003, Metcalf2005, Xu2012, Inoue2016}. Considering line-of-sight intergalactic haloes, secondary or tertiary lenses, or substructures within the primary lens along with cold dark matter haloes give a more robust explanation for microlensing in radio and infrared sources \citep{Inoue2016}. The identification of J0911+0550 lens as a galaxy cluster with complex substructures \citep{Kneib2000} is consistent with this result. 

Microlensing observations can be used to calculate the size of the broad line and continuum emission regions \citep{Vernardos2024}. Notably, the smaller the emission source in the AGN core is, the greater the magnification factor becomes \citep{Vernardos2024}. Therefore, for J1118+0745 whose radio emission might be due to AGN core based on its flat spectral energy distribution, the lack of significant quad lens component detections may indicate a strong microlensing effect on a small radio emission region.

For J0911+0550 and J1118+0745, higher spatial resolution observations in multiple wavelengths are needed to understand the exact cause of microlensing. These observations would better constrain the fluxes, identify microlensing structures and potentially explain the third significant detection in C-band for J1118+0745. 

Similarly, resolving the nature of the J0041-0203, J1010+5705 and J2246+0344 secondary detections would require either deeper radio or optical imaging to push below the current flux limit, or near-infrared spectroscopy to obtain an independent redshift for the secondary component.

The advance of telescope technologies across all bands will particularly benefit AGN studies. It will be possible to detect even older, fainter and more distant AGNs. For example, the anticipated Next Generation Very Large Array will have baselines stretching across North America, which will revolutionize the detection of radio sources. It will be able to probe an order of magnitude fainter objects with an order of magnitude higher resolution compared to the VLA \citep{Rosero2020}. The resulting fainter and more distant radio AGNs will allow us to better understand radio AGN evolution and interactions in more detail, and detect dual AGNs with much higher certainty.

\section{Acknowledgments}\label{sec:acknowledgments}
This work is supported by NSF grant AST-2108162. Y.S. acknowledges partial support from NSF grant AST-2009947. 

The National Radio Astronomy Observatory is a facility of the National Science Foundation operated under cooperative agreement by Associated Universities, Inc. 

This work has made use of data from the European Space Agency (ESA) mission {\it Gaia} (\url{https://www.cosmos.esa.int/gaia}), processed by the {\it Gaia} Data Processing and Analysis Consortium (DPAC, \url{https://www.cosmos.esa.int/web/gaia/dpac/consortium}). Funding for the DPAC has been provided by national institutions, in particular the institutions participating in the {\it Gaia} Multilateral Agreement \citep{Gaia2016, Gaia2023}.

Funding for the Sloan Digital Sky Survey V has been provided by the Alfred P. Sloan Foundation, the Heising-Simons Foundation, the National Science Foundation, and the Participating Institutions. SDSS acknowledges support and resources from the Center for High-Performance Computing at the University of Utah. SDSS telescopes are located at Apache Point Observatory, funded by the Astrophysical Research Consortium and operated by New Mexico State University, and at Las Campanas Observatory, operated by the Carnegie Institution for Science. The SDSS web site is \url{www.sdss.org}. SDSS is managed by the Astrophysical Research Consortium for the Participating Institutions of the SDSS Collaboration, including the Carnegie Institution for Science, Chilean National Time Allocation Committee (CNTAC) ratified researchers, Caltech, the Gotham Participation Group, Harvard University, Heidelberg University, The Flatiron Institute, The Johns Hopkins University, L'Ecole polytechnique f\'{e}d\'{e}rale de Lausanne (EPFL), Leibniz-Institut f\"{u}r Astrophysik Potsdam (AIP), Max-Planck-Institut f\"{u}r Astronomie (MPIA Heidelberg), Max-Planck-Institut f\"{u}r Extraterrestrische Physik (MPE), Nanjing University, National Astronomical Observatories of China (NAOC), New Mexico State University, The Ohio State University, Pennsylvania State University, Smithsonian Astrophysical Observatory, Space Telescope Science Institute (STScI), the Stellar Astrophysics Participation Group, Universidad Nacional Aut\'{o}noma de M\'{e}xico, University of Arizona, University of Colorado Boulder, University of Illinois at Urbana-Champaign, University of Toronto, University of Utah, University of Virginia, Yale University, and Yunnan University \citep{SDSS2025}.

This paper is based (in part) on results obtained with LOFAR-ERIC equipment. LOFAR \citep{LOFAR2013} is the Low Frequency Array designed and constructed by ASTRON.

This publication makes use of data products from the Wide-field Infrared Survey Explorer (WISE), which is a joint project of the University of California, Los Angeles, and the Jet Propulsion Laboratory/California Institute of Technology. WISE is funded by the National Aeronautics and Space Administration. The unWISE Catalog analysis was run on the Odyssey cluster supported by the FAS Division of Science, Research Computing Group at Harvard University, and on the National Energy Research Scientific Computing Center, a DOE Office of Science User Facility supported by the Office of Science of the U.S. Department of Energy under Contract No. DE-AC02-05CH11231. The unWISE nebulosity CNN was trained on the XStream computational resource, supported by the National Science Foundation Major Research Instrumentation program (ACI-1429830) \citep{Wright2019, Meisner2024}.

This research is based on observations made with the NASA/ESA Hubble Space Telescope obtained from the Space Telescope Science Institute, which is operated by the Association of Universities for Research in Astronomy, Inc., under NASA contract NAS 5–26555.

This work is based in part on observations made with the NASA/ESA/CSA James Webb Space Telescope. The data were obtained from the Mikulski Archive for Space Telescopes at the Space Telescope Science Institute, which is operated by the Association of Universities for Research in Astronomy, Inc., under NASA contract NAS 5-03127 for \textit{JWST}.

\textit{Software:} This research made use of the following software: Astropy \citep{Astropy2013}, APLpy \citep{APLpy2012}, CASA \citep{CASATeam2022}, Matplotlib \citep{Matplotlib2007} and SciPy \citep{Scipy2020}.

\textbf{Data Availability:} The authors confirm that the data supporting the findings of this study are available within this article. Raw SDSS and Gaia data are available publicly, raw VLA data and its products can be made available upon reasonable request to the corresponding author.

All the \textit{HST} and \textit{JWST} data used in this paper can be found in MAST: \dataset[10.17909/f7td-az53]{http://dx.doi.org/10.17909/f7td-az53}.

\bibliography{refs}
\end{CJK*}
\end{document}